\documentclass[11pt a4paper]{article}
\usepackage{jheppub}
\usepackage{bm}
\newcommand{\be}{\nopagebreak[3]\begin{equation}}
\newcommand{\ee}{\end{equation}}
\newcommand{\ba}{\nopagebreak[3]\begin{eqnarray}}
\newcommand{\ea}{\end{eqnarray}}

\newcommand{\bea}{\begin{eqnarray}}
\newcommand{\eea}{\end{eqnarray}}

\newcommand{\de}{\partial}

\def\de{\partial}

 \def\slash#1{\setbox0=\hbox{$#1$}#1\hskip-\wd0\dimen0=5pt\advance
       \dimen0 by-\ht0\advance\dimen0 by\dp0\lower0.5\dimen0\hbox
         to\wd0{\hss\sl/\/\hss}}
\def\be{\begin{equation}}

\def\ee{\end{equation}}
\def\ba{\overline }

\def\bea{\begin{eqnarray}}
\def\eea{\end{eqnarray}}

\def\EE{{\cal E}}
\def\PP{{\cal P}}

\def\7{\tilde}
\def\8{\hat}

 \def\slash#1{\setbox0=\hbox{$#1$}#1\hskip-\wd0\dimen0=5pt\advance
       \dimen0 by-\ht0\advance\dimen0 by\dp0\lower0.5\dimen0\hbox
         to\wd0{\hss\sl/\/\hss}}

\newcommand{\eq}[1]{(\ref{eq:#1})}

\def\={{\;=\;}}\def\+{{\;+\;}}
\newcommand{\bseq}{\nopagebreak[3]\begin{subequations}}
\newcommand{\eseq}{\end{subequations}\noindent}
\newcommand{\bee}{\nopagebreak[3]\begin{equation*}}
\newcommand{\eee}{\end{equation*}}

\begin{document}

\subheader{\hfill {\rm }}
\preprint{{\bf  ICCUB-21-009}} 
\title{World-Line Description of Fractons}


\author[a]{Roberto Casalbuoni}
\affiliation[a]{Department of Physics and Astronomy, University of Florence and
INFN, Florence, Italy}
\author[b]{Joaquim Gomis}\affiliation[b]{Departament de F\'{\i}sica Qu\`antica i Astrof\'{\i}sica and Institut de Ci\`encies del Cosmos (ICCUB), 
Universitat de Barcelona, Mart\'{\i} i Franqu\`es 1, E-08028 Barcelona,
 Spain}
\author[c,d,e]{Diego Hidalgo}\affiliation[c]{Centro de Estudios Cient\'ificos (CECs), Av. Arturo Prat 514, Valdivia, Chile}, \affiliation[d]{Departamento de F\'isica, Universidad de Concepci\'on, Casilla 160-C, Concepci\'on, Chile,}\affiliation[e]{ Instituto de Ciencias F\'isicas y Matem\'aticas, Universidad Austral de Chile, Casilla 567, Valdivia, Chile}.

\emailAdd{casalbuoni@fi.infn.it}\emailAdd{gomis@ecm.ub.es}\emailAdd{dhidalgo@cecs.cl} 


\begin{abstract}
{
We formulate the word-line approach of the field theory of fractons and their symmetries. The distinction between the different models is based on their dispersion relations for the energy. In order to study the sub-system symmetries, we construct the Routhians associated with the particle Lagrangians considered. 
We also construct the pseudoclassical description of spinning fractons.
}

\end{abstract}
\maketitle

\section{Introduction}

There has been recent interest in condensed matter physics to study lattice models with some peculiar properties, like the infinite degeneracy of the ground state and the existence of restricted motions along lines and/or planes. The excitations described by these models are known as ``fractons” \cite{Nandkishore:2018sel, Pretko:2020cko}. These states have zero energy but may have high momentum showing a mixing between the UV and the IR regions.
 Standard lore is that the long-distance properties of a lattice can be described in terms of field theories. It seems unlike for this kind of lattices since, in a usual field theory, the ground state is unique and low energy means to suppress the high momenta modes. In other terms, there is no UV/IR mixing.\footnote{ In non-commutative field theories, this UV/IR mixing phenomena occurs, see e.g \cite{Minwalla:1999px}.}
 However, the continuum limit of these lattice models can be formulated in terms of field theories.
 
In a series of papers \cite{Seiberg:2020bhn,Seiberg:2020wsg,Gorantla:2020xap,Gorantla:2021svj,Rudelius:2020kta} it has been shown that it is possible to consider field theories  in $d+1$ dimensions with these peculiar properties. The main idea is to consider field theories that give up the invariance under the rotation group but consider only discrete subgroups like in the lattice models.  For instance, the dispersion relation in a scalar field theory in  $d=2$, considered 
in \cite{Seiberg:2020bhn}, is given by
\be
E^2-\frac{p_x^2 p_y^2}{\mu^2}\, = \,0\,,
\ee
where $\mu$ has dimension of  a momentum.
This relation shows that there are infinite solutions with zero energy. For example, states with $p_x=0$ and $p_y\not=0$, and that these solutions have a restricted motion. In the previous example, the motion is along the spatial axis $y$. Therefore, this field theory reproduces the main features of the lattice models mentioned at the beginning. 

The fracton models that we will consider here are classified as $(m_E,n_p)$ where $m_E$ is the exponent of the energy in the dispersion relation and $n_p$ is the overall power of the 
space momenta \cite{Karch:2020yuy}.
  We will consider two classes of models: the first class is  invariant under spatial rotations of $90^0$ and depending quartically on the spatial momenta.
 In the second class, the models are invariant under rotations of $180^0$ and depend quadratically on the spatial momenta. Therefore, one goes from the first class to the second class of models via the substitution $p_i^2\to p_i$. These models will be identified as the $(m_E,n_p)^\prime$ models.

As it is well known, Feynman was the first to consider the particle approach to quantum field theory \cite{Feynman:1950ir,Feynman:1951gn}, today this  is known as the world-line approach.
In this note, we give the first steps towards constructing a world-line formulation of the field theory of fractons and their symmetries. We will construct particle models whose classical Lagrangian describe a point particle moving in space-time. 
As we shall see, using the dispersion relations of the field theory for the fractons will allow us to derive a Lagrangian  describing the classical space-time motion of these particles. The construction will be done in $2+1$ dimensions. The extensions to other dimensions can be done  using the same technique.

 An interesting aspect of the second class models is that, due to the quadratic character of the dispersion relation in the space momenta, it is possible to reexpress the kinetic term as a quadratic form in a non-Euclidean plane. The original coordinates are nothing but the light-cone coordinates of this plane. Correspondingly, the fractons can be seen as particles moving along the light-cone of the non-Euclidean plane.

We will also study all possible point symmetries of these models by constructing equations  analogous to the  Killing equations of a particle in a curved background. 
In ref. \cite{Karch:2020yuy} the sub-system symmetries of the fracton models were studied via a partial Fourier transform. In our approach, this study will be done using
 the Routhian (see, for instance, \cite{Landau:1976gn}) associated with the particle Lagrangians, which is equivalent to take a partial Legendre transform.

We will also consider the pseudoclassical world-line description of fractons by introducing Grassmann variables as was done to construct the pseudo-classical description of the Dirac equation \cite{Barducci:1976qu} \cite{Brink:1976sz}
\cite{Berezin:1976eg}.\\

The organization of the paper is as follows. In Section 2, we will construct the particle Lagrangians associated to the $(2,4)$ and $(1,4)$ models and we will study the related symmetries and sub-symmetries. The same will be done in Section 3, for the models $(2,2)'$ and $(1,2)'$. In Section 4, we will construct the pseudo-classical fractons. Finally, in Section 5 we will give some conclusions and outlook.

\section{ A class of Lagrangians quartic in the momenta }

In this section, we will construct two types of world-line Lagrangians with a mass-shell constraint quartic in the space momenta in $(2+1)$-space-time dimensions. We will consider the $(2,4)$- and the $(1,4)$ models with quadratic  and linear dependence on the energy respectively.

\subsection{The $(2,4)$-model}\label{model2,4}
This model is characterised by the mass-shell condition 
\be
E^2-\frac{p_x^2 p_y^2}{\mu^2}\, = \, 0 \,, \label{eq:1.1}
\ee
that corresponds to the world-line version of the $(2,4)$-model appearing in \cite{Seiberg:2020bhn}.\\

The canonical Lagrangian with the previous constraint is
\be\label{Lagrangian2,4}
L\, = \,-E\dot{t} +p_x\dot x+p_y\dot y+\frac \lambda 2\left (E^2-\frac{p_x^2 p_y^2}{\mu^2}\right) +\pi_\lambda\dot\lambda-\pi_\lambda\eta\,,
\ee
with the derivatives done with respect to a parameter $\tau$ which parametrises the world-line of the particle. The reason for the presence of the last two terms is as follows: in the following we would like to consider $\lambda$ as a dynamical variable and, for this, we have added the canonical term $\pi_\lambda\dot\lambda$. On the other hand, the meaning of $\lambda$ is the one of a Lagrange multiplier, which justifies the last term with $\eta$ a Lagrange multiplier.\\

The equations of motion are 
\bseq\label{eom}
\bea
&&\dot{t} = \lambda E \,, \qquad \qquad \dot x = \frac{\lambda p_x p_y^2}{\mu^2}\,, \qquad\qquad  \dot y = \frac{\lambda p_x^2 p_y}{\mu^2}\,, \\
&& \dot p_x = 0\,,\qquad \qquad \dot p_y = 0\,,\qquad \qquad \dot E =0\,, \qquad \dot\lambda=\eta \,,\qquad \dot \pi_\lambda = \frac{1}{2} \left(  E^2 -\frac{p_x^2 p_y^2}{\mu^2}\right) \,.
\eea
\eseq
 Notice that $\lambda$  is constrained by $\pi_\lambda =0$, primary constraint, which comes from the variation of $\eta$. The stability of the primary constraint gives the secondary one (\ref{eq:1.1}).
 By eliminating the momenta, we get 
\be
L\, = \,
-\frac{\dot{t}^2}{2\lambda}+\frac 3{2} \sqrt[3]{\frac{\mu^2\dot x^2\dot y^2}{\lambda}}\,.
\ee
Varying with respect to $\lambda$ we obtain its expression
\be
\lambda \, = \, \sqrt{\frac{\dot{t}^6}{\mu^2\dot x^2\dot y^2}}\,.
\ee
Then, the Lagrangian in space-time variables is given by 
\be
L \, = \, \mu\sqrt{\frac{\dot x^2\dot y^2}{\dot{t}^2}}\,.
\ee
Since all the relevant quantities are one-dimensional, the Lagrangian can also be expressed in terms of the absolute values
\be
L\,=\,\mu\left|\frac{\dot x\dot y}{\dot{t}}\right| \,.
\ee
This Lagrangian is invariant under world-line diffeomorphisms. The associated first class constraint is the mass-shell condition
\eq{1.1}. The discrete global symmetries of this Lagrangian are: spatial rotations of  $90^0$, spatial inversions, and ${t}$ inversion. 
In the next subsection, we will study the continuous symmetries.

\subsubsection{Continuous symmetries of the $(2,4)$-model}

Now we study the point continuous symmetries of this particle model. In order to do that, we will consider the most general canonical point generator 
\be\label{generator2,4}
G(t,x,y,\lambda,E,p_x,p_y,\pi_\lambda) \, = \,-E\, \xi^0(t,x,y)+  p_x \xi^x (t,x,y)+p_y \xi^y (t,x,y)+\pi_\lambda \gamma (t,x,y) +F(t,x,y)\,,
\ee
with $\xi^0, \xi^x, \xi^y,$ and $\gamma$ space-time functions to be determined. The presence of the function $F$ is  to take into account the possibility that  the Lagrangian is invariant   up to a total derivative \cite{levy1969group}. The function $G$ generates the following space-time transformations
\be
\delta x^i = \{ x^i ,G\}= \xi^i (t,x,y)\,, \qquad \qquad  \qquad  \delta t =\{ t,G\}= \xi^0(t,x,y)\,, \quad i,j=1,2\,.
\ee 
The condition that $G$ must satisfy  to generate a symmetry is to be a constant of motion, namely ${d G}/{d\tau}= 0$, implying
\be
-E(\dot{t} \partial_0 \xi^0 +\dot x_i \partial^i \xi^0) +p_i (\dot{t} \partial_0 \xi^i +\dot x_j \partial^j \xi^i) + 
\frac{\gamma }{2} \left(  E^2 -\frac{p_x^2 p_y^2}{\mu^2}\right) +\dot{t} \partial_0 F +\dot x_j \partial^j F=0 \,,
\ee
where we have used the equations of motion (\ref{eom}). Notice that we have not made use of the secondary constraint $E^2- {p_x^2 p_y^2}/{\mu^2}=0$, whereas we have used the primary one, $\pi_\lambda =0$, since it vanishes identically at the Lagrangian level.
Comparing the different  orders in powers of the space-time
momenta, we obtain the following system of partial differential equations, analogous to the Killing equations for a
relativistic particle moving in a fixed curved space-time background
\bseq
\bea
-\frac{\gamma}{2}+ \lambda (\partial_x \xi_x + \partial_y \xi_y) &=& 0 \,, \qquad -\lambda \partial_0 \xi^0 +\frac{\gamma}{2} =0\,,\\
\partial_x \xi^0 &= & 0\,,\qquad  \qquad \quad  \partial_y \xi^0  =0\,,\\
 \partial_y \xi_x & = &   0\,,\qquad  \qquad \quad   \partial_0 \xi_x = 0\,,\\
\partial_x \xi_y &= & 0\,,\qquad  \qquad \quad  \partial_0 \xi_y = 0\,,\\
\partial_i F & =& 0\,, \qquad \qquad  \quad  \partial_0 F=0\,,
\eea
\eseq
which reduces to 
\be\label{Killingequations2,4}
\lambda(\tau)\,\left(\frac{\partial \xi_x (x)}{\partial x} + \frac{\partial \xi_y (y)}{\partial y} \right)-\frac{\gamma(\tau,t)}{2}  = 0 \,, \qquad \gamma (\tau,t) = 2\lambda (\tau) \frac{\partial \xi^0 (t)}{\partial t} \,.
\ee
This system may be trivially solved, obtaining symmetries with respect to space-time dilations and    translations only
\be
 \xi^0 (t) = \chi_0 t+ a_0 \,, \qquad \xi_x(x) = \eta_0 x +a_1\,, \qquad \xi_y (y) = (\chi_0 -\eta_0 )y + a_2\,,\quad \gamma (\tau) = 2\lambda(\tau)  \chi_0\,, 
\ee
where $\chi_0,\eta_0$ and $a'$s are integration constants. The generator \eqref{generator2,4} then reads
\be
G(t,x,y,\lambda,E,p_x,p_y,\pi_\lambda) \, = \,-E\, (\chi_0 t+ a_0)+  p_x (\eta_0 x +a_1)+p_y\left( (\chi_0 -\eta_0 )y + a_2\right)+2\pi_\lambda \lambda\chi_0  \,.
\ee
Then, the space-time variables and momenta transformations read
\bseq
\bea
 \delta x & =& \eta_0x +a_1 \,, \qquad \qquad \delta y=-\eta_0 y +\chi_0 y +a_2 \,, \qquad\qquad \delta t= \chi_0 t+a_0 \,, \\
 \delta p_x &=& -\eta_0 p_x \,, \qquad \quad\quad\quad  \delta p_y =-\chi_0 p_y +\eta_0 p_y \,,\qquad \qquad \quad \delta E=-\chi_0 E\,,\\
 \delta \lambda &=& 2\chi_0 \lambda \,, \qquad \quad \qquad\delta \pi_\lambda= 2\chi_0 \pi_\lambda \,.
\eea
\eseq
The generators associated to the two dilatations, two space translations and the time translation are given by
\bea
&&P_0 = -E \,, \qquad  \qquad  \qquad \qquad \qquad  P_x = p_x \,, \qquad  \qquad   P_y=p_y\,, \qquad  \qquad \\
&& D= -tE+yp_y+2\lambda \pi_\lambda \,, \qquad  \qquad \widetilde D= xp_x-yp_y\,,
\eea
satisfying the non-vanishing commutation relations\footnote{Using the conventions for the Poisson brackets $\{t,E\}=-1\,,  \{x,p_x\}=+1\,,  \{y,p_y\}=+1\,.
$} 

\bea
&&[D,P_0] = P_0\,, \qquad  \qquad [D,P_x] =0\,,  \qquad  \qquad \quad  [D,P_y]=P_y\,,\\
&&[\widetilde D, P_0]=0\,,  \qquad  \qquad \quad [\widetilde D, P_x] =P_x\,, \qquad  \qquad [\widetilde D, P_y]=-P_y\,. 
\eea
Notice that there are no special conformal transformations. 
A fracton configuration like  $E=0, p_x=0$  has a restricted mobility,
since it can only moves in the $y$ spatial direction. The configuration preserves some of the symmetries of the Lagrangian, precisely it preserves the two dilatations and the spatial translation along the $y$ spatial direction.

On the other hand, as we have seen, the number of continuous symmetries of the world-line Lagrangian is finite-dimensional in contrast with the ansatz of \cite{Karch:2020yuy}.
In this reference, the authors consider the symmetries of two  
$(1+1)$-dimensional sub-systems obtained by doing the Fourier transform,
one in $p_x$ and the other in $p_y$, of the field theory Lagrangian. In any of these sub-systems, there is an $(1+1)$-infinite-dimensional conformal symmetry. From the world-line approach point of view, this corresponds to consider the Routh functionals, {\it i.e}, a partial Legendre 
transformation associated with the Lagrangian \cite{Landau:1976gn}.

\subsubsection{Routhian for $(2,4)$-model}

 In the Lagrangian \eqref{Lagrangian2,4} both $x$ and $y$ are cyclic coordinates. Therefore, Routh's procedure can be carried out either along the $x$ or $y$ spatial direction. Notice that the procedure used here is analogous, from a field theoretical point of view, to take a partial Fourier transform of the field Lagrangian (see \cite{Karch:2020yuy}). Let us consider the following partial Legendre transformation with respect to the $y$ coordinate given by  \be\label{Routhian2,4}
R_{(x)} \, =\,L- p_x \dot{x} \, =\,   -E \dot{t} + p_y \dot y +\frac{\lambda}{2} \left( E^2 -\frac{p_x^2 p_y^2}{\mu^2}\right)
+\pi_\lambda\dot\lambda-\pi_\lambda\eta\,,
\ee
with Hamilton's equations 
\be
\dot{p}_y = 0\,,\qquad   \dot{y} = \frac{\lambda p_x^2 p_y}{\mu^2}\,.
\ee
 Routhian functional \eqref{Routhian2,4} in configuration 
 space is given by
\be
R_{(x)}  = \frac{1}{2\lambda} \left(-\dot{t}^2 +\frac{\mu^2}{p_x^2} \dot{y}^2\right)\,.
\ee
Because $p_x =cte$, we can define $\widetilde{y}\equiv {\mu y}/{p_x}$, then
\be
 R_{(x)}  = \frac{1}{2\lambda} \left(-\dot{t}^2 + \dot{\widetilde y}^2\right)\,,
\ee
which corresponds to the Lagrangian of a massless particle in $(1+1)$-space-time dimensions. Therefore, there are infinite dimensional symmetries corresponding to the conformal group in $(1+1)$-space-time dimensions. The same discussion holds if we perform the partial Legendre transform for the $x$ coordinate.

The authors of  \cite{Karch:2020yuy} argue that the full symmetry of the $(2,4)$-model could be obtained by computing the closure of the
 infinite-dimensional conformal groups.  
 In our case, the symmetries of the world-line Lagrangian for the fracton are not the closure of the symmetries of the two conformal groups.
From our point of view, we notice that in both cases, one of the momenta is kept fixed and should not be considered a canonical variable. Therefore, the symmetry of the entire model arises by considering only the transformations of the two conformal groups that do not depend on the value of the momenta that are kept fixed. These symmetries are precisely the translations and the dilations, in agreement with the study of the Killing equations made in the previous section.

\subsection{The $(1,4)$-model}

 The $(1,4)$-model corresponds to the dispersion relation
\be
E \, = \, \frac{p_x^2 p_y^2}{\mu^3}\,,\label{eq:2.25}
\ee
and can be described by the Lagrangian
\be
L \, = \, -E \dot{t} +p_x \dot x + p_y \dot y +\lambda \left( E -\frac{p_x^2 p_y^2}{\mu^3}\right) +\pi_\lambda\dot\lambda-\pi_\lambda\eta \,,
\ee
with equations of motion given by
\bea
\dot{t} & =&\lambda \,, \qquad \dot y = \frac{2\lambda}{\mu^3} p_x^2 p_y \,, \qquad \dot x = \frac{2\lambda}{\mu^3} p_x p_y^2 \,, \\
\dot{p}_x & =& 0 \,, \qquad \dot{p}_y =0  \,, \qquad \qquad \dot{\pi}_\lambda =E -\frac{p_x^2 p_y^2}{\mu^3} \,.
\eea
Notice that $\pi_\lambda =0$ is a primary constraint, while the secondary constraint is given by (\ref{eq:2.25}). \\

The configuration space Lagrangian is given by
\be
L \, =  \, \frac{3\mu}{2} \sqrt[3]{\frac{\dot x^2\dot y^2}{2 \dot{t}}}\,.
\ee
This Lagrangian is invariant under world-line diffeomorphisms. The associated first class constraint is the mass-shell condition \eqref{eq:2.25}. The discrete global symmetries of this Lagrangian are: spatial rotations of  $90^0$, spatial inversions and 
$ x\to - x$ and $y\to - y$.

\subsubsection{Continuous symmetries of the $(1,4)$-model}

As in the previous section, we are interested in an analysis of the continuous space-time symmetries 
To this end, let us consider the following point generator transformation
\be
G(t,x,y,\lambda,E,p_x,p_y,\pi_\lambda) \, = \,-E\xi^0(t,x,y)+  p_x \xi^x (t,x,y)+p_y \xi^y (t,x,y)+\pi_\lambda \gamma (\lambda,t,x,y) \,.
\ee
Requiring that $G$ generates a symmetry, we get the system
for the unknown functions
\be
\frac{\partial \xi_x (x)}{\partial x} +\frac{\partial \xi_y (y)}{\partial y} -\frac{1}{2}\frac{\partial \xi^0 (t)}{\partial t} =0\,, \qquad \gamma = \lambda(\tau)  \frac{\partial \xi^0 (t)}{\partial t} \,.
\ee
The solution is
\be
\xi^0 (t) = c_0 +2(a_1+b_1)t \,, \qquad \xi_x (x) = b_0 +b_1 x\,, \qquad \xi_y (y) = a_0 +a_1 y\,, \qquad \gamma = 2\lambda (a_1+b_1)\,,
\ee
with $a's, b's$ and $c_0$ arbitrary constants. As wee see, the symmetries are the same of the $(2,4)$-model at Lagrangian level, but the reescaling in the dilations are different.  
\subsubsection{Routhian for $(1,4)$-model}

In order to study the sub-dimensional symmetries of this model, let us perform a partial Legendre transformation along the $y$-direction by considering the Routhian functional
\be
R_{(x)} \, = \, L-p_x \dot{x} \, = \,  -E \dot{t}  + p_y \dot y +\lambda \left( E -\frac{p_x^2 p_y^2} {\mu^3}\right)+\pi_\lambda\dot\lambda-\pi_\lambda\eta\,,
\ee
with the equations of motion
\be
\dot{t} \, = \, \lambda \,, \qquad \dot{y} =\lambda \frac{2p_x^2 p_y}{\mu^3}\,,\qquad p_x =cte\,, \quad \dot{E} = 0\,,\quad  \dot\pi_\lambda=E -\frac{p_x^2 p_y^2}{\mu^3}\,.
\ee
The Routhian in configuration space is
\be
R_{(x)}= \frac{m}{2}\frac{\dot y^2}{\dot t}\,,
\ee
that is the Lagrangian for a one-dimensional non-relativistic particle of mass $m\equiv  {\mu^3}/{2p_x^2 }.$ For studying the symmetries of this model, we consider the following generator
\be
G(t,y,E,p_y,\pi_\lambda) \, = \,-E\xi^0(t,y)+p_y \xi^y (t,y)+
 \gamma(t,y) \pi_\lambda +F(t,y) \,.
\ee
 By requiring $G$ be a generator of symmetries, we get the following ordinary differential equations
\bseq
\bea
-\gamma  +2 \partial_y \xi^y &= &0\,, \qquad \qquad \qquad  \qquad   \gamma   = \partial_0 \xi^0\,,\\
\partial_y \xi^0 & =&0\,,\qquad \partial_0 \xi^y + \frac{2p_x^2 }{\mu^3}\,\partial_y F =  0\,,\qquad \partial_0 F = 0\,,
\eea
\eseq
whose solution is given by
\be
\xi^0 (t) =  b_0 +2a_1 t + a_3 t^2 \,, \quad \xi_y (t,y)\, = \, a_0 + a_2 t+a_1 y +a_3 ty \,, \quad F(y) \, = \,-m \, \left(a_2 y + \frac{a_3}{2} y^2\right)\,,
\ee
with $a's$ and $b's$ arbitrary constants. Then, the generator reads
\be
G(E,p_y,t,y) \, = \, -E\left(  b_0 +2a_1 t + a_3 t^2 \right) + p_y  \left( a_0 + a_2 t+a_1 y +a_3 ty \right)-m  \left(a_2 y + \frac{a_3}{2} y^2\right)\,.
\ee
 From here, we read off the generators
\bea
H &=& -E\,,\qquad \qquad   P =  p_y\,,\qquad \qquad \quad  D  = p_y y- 2Et   \,,\\
G&= & p_y t- my\,, \qquad C=  -Et^2 + p_y t y - \frac{m}{2} y^2\,.
\eea
The non-vanishing commutators are given by
\bseq
\bea	
&& [D,H] \, = \, 2 H\,, \qquad [D,P] \, = \, P \,, \qquad [C,P] \, =\, G\,,\qquad [D,G] \, = \,-G\,,\\
&& [D,C] \, = \,-2C\,,\qquad [H,C] \, = \, -D\,, \qquad [H,G] = -P \,, \qquad [G,P]=-m\,.  
\eea
\eseq	
This is the Schr\"odinger algebra in $(1+1)$-space-time dimensions. When we consider the Routhian associated to the coordinate $y$ we obtain the same symmetries. Notice the sub-system symmetries of the $(1,4)$-model are different from the $(2,4)$-model. 

\section{ A class of Lagrangians quadratic in the space momenta}

In this section, we will construct two types of world-line Lagrangians with a 
mass-shell constraint quadratic in the space momenta in $(2+1)$-space-time dimensions. 
{

\subsection{ The $(2,2)'$-model}
 In this case, the dispersion relation  is
\be
E^2\, = \, p_x p_y \,.
\ee
 Introducing the three vector
$
p^\mu= (E, p_x,p_y)\,
$  
and the metric tensor
\be
g_{\mu\nu}=\left(
 \begin{matrix}
      -1 & 0& 0 \\
      0 & 0 & 1/2 \\
      0&1/2&0\\
   \end{matrix}\right)\,,
\ee
the dispersion relation can be rewritten in the form
\be
 p^\mu g_{\mu\nu} p^\nu\, = \, 0 \,.
 \ee
 
Now, introducing the space momenta
 \be\label{q-variables}
 q^1= \frac{1}{2}\left(p_x+p_y\right)\,,~~~ ~~ q^2=\frac{1}{2} \left(p_x-p_y\right)\,,
 \ee
 the mass-shell constraint becomes
 \be
 q^\mu\eta_{\mu\nu}q^\nu=(-q_0^2+q_1^2-q_2^2)=0\,,~~~\mu,\nu=0,1,2\,,\label{eq:3.5}
 \ee
 where $q^\mu=(E,q^1,q^2)$ and $\eta_{\mu\nu}=(-1,1,-1)$. Therefore the mass-shell condition for this case, can be thought as describing a massless particle in a non-Euclidean space with signature $(1,-1)$. The original momenta are  the light-cone momenta in this space. From this point of view, a fracton is a particle moving along the light-cone of a non-Euclidean space.\\
  
 The dispersion relation is invariant under the following transformation
 \be
 \delta E=\frac 12\epsilon\,  p_x\,, \qquad \delta p_x=0\,,\qquad \delta p_y =\epsilon\, E\,.
 \ee
 A fracton configuration with  $E=p_x=0$ is also invariant. This describes a movement along the $y$ spatial direction.  The space-time configuration Lagrangian can be written in terms of the original space-time variables as
 \be
L=\frac1{2\lambda}\dot x^\mu g_{\mu\nu}\dot x^\nu =\frac{1}{2\lambda}\left(-\dot{t}^2+\dot x\dot y\right)\,,
\ee
or if we introduce a new space coordinates 
\be
y_1= \frac{1}{2}(x+y)\,, \qquad  y_2 =\frac{1}{2} (x-y)\,,
\ee
we find
\be
L= \frac1{2\lambda}\dot y^\mu \eta_{\mu\nu}\dot y^\nu =\frac 1{2\lambda}\left(-\dot{t}^2+\dot y_1^2-\dot y_2^2\right)\,.
\ee
Therefore, this Lagrangian describes a massless particle in $(1+2)$-space-time-dimensions. It follows that $L$ is invariant under the conformal group  $O(3,2)$. The space-time symmetries in this case are finite dimensional.

\subsubsection{Continuous symmetries of the $(2,2)'$-model}

In this case the generator of the point continuous symmetries is 
\be
G(x^\mu,\lambda,p_\mu,\pi_\lambda) \, = \,  p_\mu \xi^\mu (t,x,y) + \pi_\lambda \gamma (\tau,t,x)  \,,\quad \mu =0,1,2\,,
\ee
and the Killing equations are
\be
 \partial^\mu \xi^\nu + \partial^\nu \xi^\mu= \eta (\tau) g^{\mu \nu}\,, \qquad \gamma =\lambda(\tau) \eta(\tau)\,,
\ee
whose solutions give the conformal group in $(2+1)$-space-time dimensions.

\subsubsection{Routhian for $(2,2)'$-model}

Let us consider the following Routhian functional along the $y$ spatial direction
\label{Routhian12,2} 
\be
R_{(x)}\, = \, L- p_x \dot x = -E\dot{t} +p_y \dot{y} +\frac{\lambda}{2}\left( E^2-p_xp_y\right)
+\pi_\lambda\dot\lambda-\pi_\lambda\eta\,.
\ee
The Euler-Lagrange equations are 
\be
\dot{t}=\lambda E\,, \qquad \dot{y} \, =  \frac{\lambda}{2} p_x\,, \quad \dot{p}_y=0\,, \quad \dot{E}=0\,
\qquad \dot\pi_\lambda=\frac{1}{2} \left(E^2-p_xp_y \right)\,.
\ee

 The Routhian  in configuration space is given by
 \be
 R_{(x)}=-\frac 14 \frac{\dot t^2}{\dot y} p_x\,.
 \ee
 This expression looks rather peculiar but it can be understood by looking at the Routhian in coordinate space for the case $(1,4)$. Except for trivial factors, the two Routhians can be obtained one from the other by exchanging time with the space coordinate $y$. This is due to the fact that in the case of the Routhian for the $(1,4)$-model, describing a non-relativistic one-dimensional particle, the effective dispersion relation is of type $(1,2)$, whereas in the actual case is of type $(2,1)$, with an exchange of  the energy with the spatial momentum.\\
 
By studying the Killing equations with the generator
\be
G(t,y,E,p_y,\pi_\lambda) \, = \,-E\xi^0(t,y)+  p_y \xi^y (t,y)+ \pi_\lambda\gamma(t,y)
+F(t,y)\,,
\ee
and considering only primary first class constraints, we now find the system
\bseq
\bea
\gamma+\partial_0 \xi^0  -\frac{1}{2} \partial_y \xi_y  &= &0 \,,\qquad \partial_0 \xi^y=0\,,\qquad \gamma=0\,,\\
 \partial_y \xi^0 - \frac{2}{p_x} \, \partial_0 F&=&0\,,\qquad \partial_y F =  0\,,
\eea
\eseq
whose solution is given by
\be
\xi^0(t,y) \, = \, b_0 + a_1 t + b_2\, y + a_2 \, t\,y \,,\qquad \xi_y(y) = a_0 + 2 a_1 y + a_2  y^2 \,,\qquad F(t)\, =\, \frac{p_x}{2} \left(b_2 \, t+ \frac{a_2 }{2}\, t^2\right)\,.
\ee
Then, the generators yields
\be
G= -E \left(b_0 + a_1 t + b_2\, y + a_2 \, t\,y \right) +  p_y \left(a_0 + 2 a_1 y + a_2  y^2 \right) +\frac{p_x}{2} \left(b_2 \, t+ \frac{a_2 }{2}\, t^2\right)\,,
\ee
where $a's$ and $b's$ are arbitrary constants. The generators of this symmetry (defining $p_x \equiv 2m$) read
\bseq
\bea
P \, & =&  \, p_y \,,\qquad D =  2 p_y\, y -  E\,t\,, \qquad H =  -E\,, \\ 
   C& =& \frac{m}{2}\, t^2 - E\, t \,y + p_y \, y^2\,,\qquad G  =   m\, t  -E\, y \,.
\eea
\eseq
The non-vanishing commutators are given by
\bea
&& [D,H] = H\,, \qquad [D,P] = 2P\,, \qquad [C,P]=D\,, \qquad [D,G] =-G\,,\\
&& [D,C] = -2C \,, \qquad [C,H] =G  \,, \qquad [G,P] = H \,,\qquad  [G,H] =m\,.
\eea
Notice that the last two commutators correspond to the ones appearing in Carroll algebra in two dimensions with a central charge (see for instance \cite{Gomis:2020wxp}). We shall call the previous algebra the $(1+1)-$Schr\"odinger Carroll algebra.

\subsection{The $(1,2)'$-model}

The dispersion relation of this model   is
\be
E\, = \,\frac 1\mu  p_x p_y\,.
\ee
In terms of the $q$ variables \eqref{q-variables}, it yields
\be
E= \frac 1\mu \left(q_1^2-q_2^2 \right)\,,
\ee
which is the dispersion relation for a non-relativistic particle of mass $\mu/2$ in a non-Euclidean space. This model can be treated as the previous ones. Starting from
\be
L\, = \, -E\dot{t} +p_i\dot x^i+ \lambda \left (E-\frac 1\mu p_x p_y \right)+\pi_\lambda\dot\lambda-\pi_\lambda\,,\qquad i,j=1,2\,.
\ee
Proceeding as before we find
\be
L=\mu\frac{ \dot x\dot y}{\dot{t}}\,.
\ee
This Lagrangian is similar to the one of the $(2,4)$-model of Section \ref{model2,4}. The main difference is in the signs. In fact $\sqrt{\dot x^2}=|\dot x|$ and not equal to $\dot x$. This implies different properties in the symmetries. The Lagrangian  model is not invariant under rotations of $90^0$, but it is invariant under rotations of $180^0$. 

\subsection{Continuous symmetries of the $(1,2)'$-model}

To study global symmetries of this model, let us consider the generator 
\be
G(t,x,y,E,p_x,p_y,\pi_\lambda) \, = \, -E\xi^0(t,x,y)+  p_x \xi^x (t,x,y)+ p_y \xi^y (t,x,y)+\pi_\lambda \gamma (t,x,y)+F(t,x,y)\,.
\ee
For $G$ to generate a symmetry, we find the system
\bseq
\bea
\mu \, \frac{\partial \xi_x (t,x)}{\partial t }  + \frac{\partial F(x,y)}{\partial y}  & =& 0\,, \quad \mu \, \frac{\partial \xi_y (t,x)}{\partial t }  + \frac{\partial F(x,y)}{\partial x}   = 0\,,\\
 \frac{\partial \xi_x (t,x)}{\partial x }  + \frac{\partial \xi_y (t,y)}{\partial y } -  \frac{\partial \xi^0 (t)}{\partial t }  & =& 0\,, \quad \gamma (t)  = \frac{\partial \xi^0 (t)}{\partial t }\,.
\eea
\eseq
The solution of this system is
\bea
\xi^0 (t) & =& a_0 + 2b_1 t+ a_2 t^2\,,\\
\xi_x (t,x) & =& b_0 + b_2 t + b_1 x + a_2 t x\,,\\
\xi_y (t,y) & =& c_0 +c_2 t +b_1 y + a_2 t y\,,\\
F(x,y) & =& -\mu\, \left( b_2 x +c_2y +\frac{a_2}{2} \left(x^2 +y^2\right)   \right)\,, \\
\gamma (t) & =& 2(b_1 + a_2\,t)\,.
\eea
with $a's, b's$ and $c's$ again arbitrary constants. Then, the generator reads
\bea
\nonumber  G(t,x,y,E,p_x,p_y,\pi_\lambda)  &=&  -E(a_0 + 2b_1 t+ a_2 t^2)+  p_x (b_0 + b_2 t + b_1 x + a_2 t x)+ p_y (c_0 +c_2 t +b_1 y + a_2 t y )\\
&&  +\,2\pi_\lambda  (b_1 + a_2\,t)-\mu\, \left( b_2 x +c_2y +\frac{a_2}{2} \left(x^2 +y^2 \right)   \right)\,.
\eea
From here, we get the following generators
\bseq
\bea
H & =& -E\,, \qquad  \qquad P_i  = p_i\,,\qquad\qquad G_i = p_i t - \mu \delta_{ij}x^j \,,\\
C& =&-Et^2+ t \,p_i x^i  - \frac{\mu}{2} \delta_{ij}x^i x^j +2t \pi_\lambda\,, \qquad D = p_i x^i  -2 Et + 2\pi_\lambda\,.
\eea
\eseq
The generators satisfy the following non-vanishing commutators\bea
\nonumber && [D, H] = 2H \,, \qquad [D,P_i] = P_i \,, \qquad [C,G_i]=G_i\,, \qquad [D,G_i] =-G_i \,,\\ 
&& [D,C] =-2C\,,\qquad [C,H]=D \,, \qquad [G_i,P_j] = - \mu \delta_{ij} \,, \qquad[G_i,H] = P_i\,,
\eea
with $i,j=1,2$. This algebra corresponds to the $(1+2)-$Schr\"odinger algebra.

\subsubsection{The Routhian for $(1,2)'$-model}
The Routhian along the $y$ spatial direction is given by
\be
R_{(x)}=-E\dot{t} +p_y\dot y+ \lambda \left (E-\frac 1\mu p_x p_y \right).
\ee
The relevant equations of motion are
\be
\dot{t}= \lambda\,, ~~~~ ~~  \dot y= \lambda \frac{p_x}\mu\,.
\ee
It results impossible to invert the momenta with the velocities. Proceeding as we did for the Routhian of the $(2,4)$-model, we define $\widetilde{y}\equiv \mu/p_x$, from which
\be
\dot{\widetilde{y}}= \dot{t} \,.
\ee
We see that this model is a branch of the Routhian case for the $(2,4)$-model. In fact, there we got a one-dimensional massless particle that can move along the two branches of the light cone, whereas in this case only one branch it is allowed. The symmetries of this case are translations and dilations.

\section{Pseudo-Classical Spinning Fractons}
\subsection{The spinning $(2,4)$ model}
In this section, we will consider the pseudo-classical spinning
fracton associated to the $(2,4)$-model, for the other models the same procedure can de done
in a similar way. The idea is considering the square root of the mass-shell constraint
\be
\Phi \,  = \, E^2-\frac{p_x^2 p_y^2}{\mu^2}\,.
\ee
Now let us introduce the Grassmann variables $\lambda^0,\lambda^1$ with the non-vanishing Dirac brackets
\be 
\{\lambda^0,\lambda^0\}^*=-i\,,\qquad \{\lambda^1,\lambda^1\}^*=i\,.
\ee
The odd constraint reads
\be
\Phi_D=-E\lambda^0+ \frac 1\mu\lambda^1 p_xp_y\,,
\ee
which verifies
\be
\{\Phi_D,\Phi_D\}^*=-i\Phi\,.
\ee
The canonical Lagrangian of the pseudo-classical spinning fracton is
\be
L_C=-E\dot t+p_x\dot x+p_y\dot y+\frac i2 \lambda^0\dot\lambda^0
-\frac i2 \lambda^1\dot\lambda^1
-\frac e2\Phi-i\chi\Phi_D\,,
\ee
and one can check that the brackets assumed for the Grassmann variables are the Dirac brackets that are obtained from this Lagrangian.

 The continuous symmetries are the same as the bosonic counterpart except that under dilations the odd Lagrange multiplier should be scaled to compensate the overall scaling factor of the odd constraint, leaving invariant the Grassmann variables. The $90^0$ rotation is still a symmetry nut the Grassmann variables must transform as follows
\be
\lambda^0\to \lambda^0, \quad\quad\lambda^1\to-\lambda^1.
\ee

At quantum level \cite{Casalbuoni:1975bj} we can realise the Grassmann in terms of the Pauli matrices
\be
\hat\lambda^0=\frac{i}{\sqrt{2}}\sigma_3\,,\qquad \hat\lambda^1=\frac{i}{\sqrt{2}}\sigma_1\,,
\ee
Then, the fracton Dirac field equation is 
\be
\left(\sigma_3 i\partial_t-\frac 1\mu\sigma_1\partial_x\partial_y\right)\Psi(t,x,y)=0\,.
\ee
The field $\Psi(t,x,y)$ also verifies the analog of the Klein-Gordon equation
\be
\left(\partial^2_t+\frac 1{\mu^2}\partial_x^2\partial_y^2\right)\Psi(t,x,y) \, = \, 0\,.
\ee
Notice that the factorization of the quartic term is not unique and it  could be done also in terms of $p_x^2$ and $p_y^2$  instead of $p_x p_y$. In this case we should have made use of three Grassmann variables instead of two and the resulting Dirac equation should look differently, more like the one that we have written in the following subsection for the $(2,2)'$-model.

\subsection{The spinning $(2,2)'$-model}

The spinning $(2,2)'$-model could  be treated in a similar way to the $(2,4)$-model. However, we will make use of the covariant form with a non-Euclidean two-dimensional space, where the dispersion relation assumes the form (see Eq.~\eqref{eq:3.5})
\be
\Phi=q^\mu\eta_{\mu\nu} q^\nu=0,~~~\, \mu,\nu=0,1,2\,,~~~\eta_{\mu\nu}=(-1,1,-1)\,.
\ee
Let us introduce a set of Grassmann variables, $\xi^\mu$, with Dirac brackets given by
\be
\{\xi^\mu,\xi^\nu\}^*=-i\eta^{\mu\nu}\,.
\ee
Then, defining
\be
\Phi_D= q^\mu\xi_\mu\,,
\ee
we get
\be
\{\Phi_D,\Phi_D\}^*= -i\Phi\,.
\ee
We can easily write a Lagragian in phase space by using the constraints $\Phi$ and $\Phi_D$. This is given by
\be
L= q^\mu \dot y_\mu+\frac i2 \xi^\mu\dot\xi_\mu -\frac e 2 \Phi -i\chi\Phi_D\,,
\ee
where $e$ and $\chi$ are Lagrange multipliers, with $\chi$ a Grassmann variable (see refs. \cite{Barducci:1976qu,Brink:1976sz,Casalbuoni:2008iy}). It is easy to show that this Lagrangian gives rise to the previous Dirac brackets  \cite{Casalbuoni:2008iy}.\\

The quantization of the Grassmann algebra leads to the Clifford algebra \cite{Casalbuoni:1975bj}
\be
[\widetilde\gamma_\mu,\widetilde\gamma_\nu]_+=\eta_{\mu\nu}\,.
\ee
The $\widetilde\gamma^\mu$ can be expressed in terms of the usual $\gamma_\mu$ as follows
\be
\widetilde\gamma^0=\frac 1{\sqrt{2}}\gamma^0\,,~~~~\widetilde\gamma^1=\frac 1{\sqrt{2}}\gamma^1\,, ~~~~\widetilde\gamma^2=\frac i{\sqrt{2}}\gamma^2\,.
\ee
Then, the corresponding wave equation is
\be
-i\widetilde\gamma^\mu\de_\mu\psi=0\,.
\ee

 \section{Conclusions an outlook}
 
We have started the construction of the world-line approach to a field theory of fractons. We have analyzed particle Lagrangians whose mass-shell constraints give the dispersion relation of $(2,4), (2,2)$ models in $(2+1)$-dimensions. The extension to other models and directions follow the same lines.

We have also studied the symmetries of the respective Lagrangians by writing the analog of Killing equations of a relativistic particle in a curved background. The construction of the associated Routhians to the Lagrangians allows to find the sub-system symmetries.

We have also initiated the construction of pseudo-classical spinning fractons that lead in a natural way to the Dirac fracton equations.

As a further study to consider, is the study of three-dimensional world-line models. Here the number of possible models increases with the number of invariants. An interesting case would be the model $(2,2)'$ in three dimensions that, in analogy with the two-dimensional case, can be seen as a massless theory in a space-time with signature $(-1,1,-1,-1)$. We hope to return to this study in the future. 

In the future we would like to consider the self-interaction of fractons by using the geometrical interaction, in the ordinary relativistic particle case, see for example \cite{Casalbuoni:1974pj}.
 The analysis will be done by using the path integral formulation.

Another interesting development will be to study the dynamics and symmetries of fractonic strings. The canonical world-sheet action for the $(2,4)$ string  with space-time coordinates $(t(\tau,\sigma),\,
x^i(\tau,\sigma), i=1,2)$, and energy and momentum densities $\EE(\sigma,\tau)$ and   $\PP_i(\sigma,\tau), i=1,2$
is given by
\be\label{Lagrangian2,4string}
S\, = \int d\tau\,d\sigma \big(-\EE\dot{t} +\PP_i\dot x^i-\frac \lambda 2\left (\EE^2-\frac{\PP_x^2 \PP_y^2}{\mu^2}\right) -\rho(-\EE t'+\PP_i {x'}^i)
\big)\,,
\ee
where dot means partial with respect to $\tau$ and prime derivate with respect to $\sigma$ with $0\le\sigma\le\pi$. The Lagrange multipliers
$\lambda$ and $\rho$ impose the first class constraints of two-dimensional world-sheet diffeomorphisms.\\

The $(2,2)^\prime$ string has the canonical action
\be\label{Lagrangian1,4string}
\, = \int d\tau\,d\sigma \big(-\EE\dot{t} +\PP_i\dot x^i-\frac \lambda 2\left (\EE^2-\PP_x \PP_y\right) -\rho(-\EE t'+\PP_i {x'}^i)\big)\,.
\ee
The study of these strings will be done in a subsequent work \cite{future}.

\section*{Acknowlegdments}
We are grateful to Jaume Gomis for careful reading and enlightening comments. The work of JG has been supported in part by MINECO FPA2016-76005-C2-1-P 
and PID2019-105614GB-C21 and from the State Agency for Research of the
Spanish Ministry of Science and Innovation through the Unit of Excellence
Maria de Maeztu 2020-203 award to the Institute of Cosmos Sciences
(CEX2019-000918-M). The Centro de Estudios Cient\'{\i}ficos (CECs) is funded by the Chilean
Government through the Centers of Excellence Base Financing Program of ANID.

\end{document}